\documentclass[twocolumn,amsmath,showpacs,amssymb,prb]{revtex4}

\usepackage{graphicx}
\usepackage{dcolumn}
\usepackage{bm}

\newcommand{\bs}{\boldsymbol}

\newcommand{\mb}{\mathbf}

\begin{document}

\preprint{APS/123-QED}

\title{Energy and momentum entanglement in parametric downconversion}

\author{Pablo L. Saldanha}\email{saldanha@df.ufpe.br}
\affiliation{Departamento de F\'isica, Universidade Federal de Pernambuco,
50670-901, Recife, PE, Brazil}

\author{C. H. Monken}
\affiliation{Departamento de F\'isica, Universidade Federal de Minas
Gerais, Caixa Postal 702, 30161-970, Belo Horizonte, MG,
Brazil}



\begin{abstract}
We present a simple treatment of the phenomenon of spontaneous parametric
downconversion consisting of the coherent scattering of a single pump photon into an
entangled photon pair inside a nonlinear crystal. The energy and momentum
entanglement of the quantum state of the generated twin photons are seen as a
consequence of the fundamental indistinguishability of the time and the position in
which the photon pair is created inside the crystal. We also discuss some
consequences of the photon entanglement.
\end{abstract}


\maketitle

\section{Introduction}

The generation of twin photons inside a nonlinear crystal by the phenomenon of
spontaneous parametric downconversion (SPDC) has been extremely useful for studying
fundamental aspects of quantum mechanics and to physically implement quantum
information protocols in recent decades. This usefulness is due to the photon pair
being entangled in many degrees of freedom, including energy,\cite{rarity90}
momentum,\cite{howell04} angular momentum,\cite{mair01} and
polarization.\cite{kwiat95} For a pair of particles in an entangled state Bell's
theorem states that we cannot associate an objective reality to each particle, and
this has profound consequences for the way we view nature.\cite{peres,mermim85}
Entanglement is also useful in quantum information science, permitting the execution
of quantum algorithms that are more efficient and secure than classical
algorithms.\cite{nielsen}

In this paper we derive an expression for the quantum state of the twin photons
generated in the process of SPDC using a simplified version of the example presented
in our recent work,\cite{saldanha11} in which the interaction between light and
matter is treated using the Bialinicki-Birula--Sipe photon wave function
formalism.\cite{birula94,sipe95} (For didactical discussions of the photon wave
function formalism, see Refs.~\onlinecite{raymer05} and \onlinecite{birula06}.) The
reader interested in a more formal treatment should consult Ref.
\onlinecite{saldanha11}.  The main idea is that when a single photon is converted
into two photons inside a nonlinear crystal, there is a fundamental uncertainty in
the location and the time when the photon pair is generated. This fact requires that
we coherently sum all possible probability amplitudes for this event. The
interference of amplitudes for generation at different times leads to energy
entanglement while the interference of amplitudes for generation at different
positions leads to momentum entanglement.

The remainder of this paper is organized as follows.  In Sec.~\ref{sec2} we discuss
the physical arguments of our approach and arrive at a general expression for
calculating the twin-photon wave function. In Sec.~\ref{sec3} we compute the
twin-photon wave function using reasonable approximations, and in Sec.~\ref{sec4} we
discuss some consequences of the system entanglement. Finally, in Sec.~\ref{sec5} we
present some concluding remarks. 

\section{Parametric downconversion as a scattering phenomenon}\label{sec2}

If an electromagnetic wave interacts with a small transparent object that responds
linearly to the electric field, such that the induced electric dipole moment of the
object is proportional to the incident electric field, the resultant scattered
electromagnetic field is a superposition of the incident field with the field
generated by the oscillating dipole moment of the object. The scattered field will
in general have a diffraction pattern from which we can deduce some properties of
the scatterer. If the incident electromagnetic field is composed by only one photon
and we use a quantum language, we can say that there is a probability amplitude for
the incident photon to be instantaneously absorbed and re-emitted by the object that
must be coherently summed with the probability amplitude for the photon to pass
without any interaction with the object in order to compute the wave function of the
scattered photon. 
The state of the transparent object  before and after the interaction with the
photon is the same, such that no correlation between the state of the object and the
state of the photon appears.

This instantaneous absorption and re-emission of a photon can be interpreted as the
absorption of the incident photon inducing the object to have an oscillating dipole
moment and this oscillating dipole moment creating the scattered photon.  For
instance, the propagation of an electromagnetic wave through a transparent linear
medium can be seen as a superposition of the incident wave being transmitted
directly through the medium with the waves generated by the oscillating charges in
the medium.\cite{feynman2,james92} 
Each component of the resultant wave propagates at the
speed of light in vacuum, but the coherent superposition of these components 
causes the resultant wave to propagate with a reduced
velocity.

In the phenomenon of parametric downconversion, light interacts with a nonlinear
medium whose nonlinear polarization is proportional to the square of the incident
electric field (or to the product of different components of the electric field). A
nonlinear scatterer like this can absorb two photons and emit one. We can interpret the phenomenon as the induced
nonlinear dipole oscillations creating one photon, but these oscillations occurring at
the expense of the absorption of two photons from the incident field. 
This situation corresponds to
the phenomenon of second-harmonic generation. By symmetry it must also be possible
that the nonlinear scatterer does the opposite---emits two photons while absorbing
one.  Such a process corresponds to parametric downconversion. 

If we have a medium with a small nonlinear scatterer placed at a fixed position
$\mb{r}'$, the interaction between an incident electromagnetic field and this
nonlinear scattering element leads to a probability amplitude for one incident pump
photon to be converted into two photons during its passage through the medium. This
situation is illustrated in Fig. 1, where photon 1 is observed at position
$\mb{r}_1$ and photon 2 at position $\mb{r}_2$, both at time $t$. Of course, there
is also a probability amplitude that no downconversion occurs, which is what happens
in most cases, but from now on we will assume we are dealing with the case when a
photon pair is created.

We assume that the incident photon has a wave function $\psi_p(\mb{r}',t')$ in the
nonlinear scatterer element and that the downconversion occurs at time $t'$, such
that the probability amplitude of generating the twin photons at time $t'$ is
proportional to $\psi_p(\mb{r}',t')$. If we assume that both photons are created at
a precise position $\mb{r}'$ and time $t'$, then according to the uncertainty
relations there must be a large uncertainty in their momenta $\hbar \mb{k}_i$ and
energies $\hbar\omega_i=\hbar{k}_i c/n_i$. Here, $\mb{k}_i$ is the wavevector,
$\omega_i$ the angular frequency, and $n_i$ the refractive index for photon $i$
($=\{1,2\}$) in the propagation medium, $k_i=|\mb{k}_i|$, $\hbar$ is Planck's
constant divided by $2\pi$, and $c$ is the speed of light in vacuum. Thus, to
compute the total probability amplitude for finding photons 1 and 2 at positions
$\mb{r}_1$ and $\mb{r}_2$ at time $t$, we must coherently add the probability
amplitudes for the propagation of photon 1 from $(\mb{r}',t')$ to $(\mb{r}_1,t)$
with all possible wavevectors $\mb{k}_1$ and angular frequencies
$\omega_1={k}_1c/n_1$, and similarly for photon 2.

Now, when propagating from  $(\mb{r}',t')$ to $(\mb{r}_i,t)$, a photon in a
plane-wave mode with wavevector $\mb{k}_i$ and angular frequency $\omega_i$
accumulates a phase $\phi=\mb{k}_i\cdot(\mb{r}_i-\mb{r}')-\omega_i(t-t')$ in the
probability amplitude. So we can write the total probability amplitude of finding
the generated photons in the positions $\mb{r}_1$ and $\mb{r}_2$ at time $t$
as\footnote{We could use spherical-wave modes to describe the propagation of photons
1 and 2 from $(\mb{r}',t')$ to $(\mb{r}_1,t)$ and $(\mb{r}_2,t)$ in Eq.~(\ref{Ac}).
This would be more precise, but the plane-wave decomposition that we use here, in
which all wavevectors appear with the same probability amplitude, leads to simpler
calculations with essentially the same results.}
\begin{eqnarray}
\label{Ac}
&&A_c(\mb{r}_1,\mb{r}_2,t;\mb{r}',t') \propto \psi_p(\mb{r}',t') \int d^3k_1 \int
d^3k_2  \nonumber\\
&&\quad \times\;\mathrm{e}^{i\mb{k}_1\cdot(\mb{r}_1-\mb{r}')-i\omega_1(t-t')}
\mathrm{e}^{i\mb{k}_2\cdot(\mb{r}_2-\mb{r}')-i\omega_2(t-t')}.
\end{eqnarray}
Here and elsewhere in this paper, unless explicitly stated the integration is
carried out over the entire domain of the integration variables.

In the above probability amplitude, we are coherently superimposing possibilities
where the energy and momentum of photons 1 and 2 have a sum that is different from
the energy and momentum of the incident photon, thus being possibilities that do not
conserve energy or momentum. This treatment is closely related to the Feynman
path-integral formalism, and the conservation of energy and momentum in the process
appears somewhat surprisingly as a consequence of the coherent sum of all
possibilities. The point is that possibilities that do not conserve say, energy,
will interfere destructively among themselves, as will become clearer in the
remainder of the paper.

\begin{figure}\begin{center}
  \includegraphics[width=7cm]{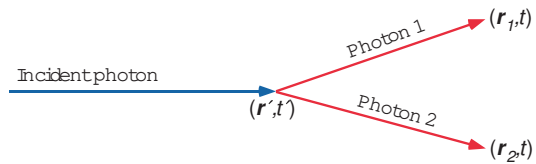}\\
  \caption{Illustration of the nonlinear scattering of one photon into two by a
scatterer at position $\mb{r}'$ at time $t'$. Photon 1 is observed at position
$\mb{r}_1$ and photon 2 at position $\mb{r}_2$, both at time $t$.}\label{fig1}
 \end{center}\end{figure}

Let us now consider that instead of a single nonlinear scatterer we have a
continuous set of nonlinear scatterers forming a nonlinear crystal, as depicted in
Fig. 2. For simplicity, we will assume that the linear response of the medium
outside the crystal is the same as inside, but the nonlinear response is absent. 
In this situation, there is a fundamental quantum indistinguishability of the
position and the time in which the photon pair is created in the process of
parametric downconversion. This condition forces us to coherently sum the
probability amplitudes of generating the photons in any part of the crystal and at
any time to find the total probability amplitude of finding photons 1 and 2 at the
positions $\mb{r}_1$ and $\mb{r}_2$ at time $t$. Figure 2 illustrates the
probability amplitudes for generating the photon pair at the spacetime points
$(\mb{r}_a,t_a)$ and  $(\mb{r}_b,t_b)$. The   twin-photon wave function
$\psi^{(2)}(\mb{r}_1,\mb{r}_2,t)$ is proportional to this total probability
amplitude, and we can write
\begin{equation}
\label{scatt}
\psi^{(2)}(\mb{r}_1,\mb{r}_2,t)\propto \int_\mathcal{V} d^3r' \int_{-\infty}^t dt'
A_c(\mb{r}_1,\mb{r}_2,t;\mb{r}',t'),
\end{equation}
where the spatial integral is taken over the crystal volume $\mathcal{V}$, and
$A_c(\mb{r}_1,\mb{r}_2,t;\mb{r}',t')$ is given by Eq. (\ref{Ac}).

\begin{figure}\begin{center}
  \includegraphics[width=7cm]{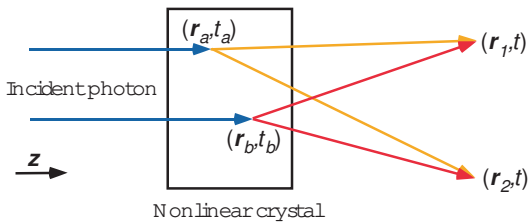}\\
  \caption{Illustration of the nonlinear scattering of one photon into two by a
nonlinear crystal. Due to the fundamental indistinguishability of the position and
time of the downconversion process, all possibilities must be coherently added. In
the figure the probability amplitudes of the conversion into the spacetime points
$(\mb{r}_a,t_a)$ and  $(\mb{r}_b,t_b)$ are represented.}\label{fig2}
 \end{center}\end{figure}

Equations~(\ref{Ac}) and (\ref{scatt}) are the essential equations of our treatment
so we discuss their physical meaning once again.  The twin-photon wave function is
obtained from the coherent scattering of a single pump photon into a photon pair
inside the nonlinear crystal.  Because there is a fundamental indistinguishability
of the position and time in which the photon pair is created, the probability
amplitudes for generation at different times and positions must be coherently added,
leading to the volume and time integrals in Eq.~(\ref{scatt}).  The probability
amplitude for a precise creation time and position is given by Eq.~(\ref{Ac}), which
takes into account the fact that these photons must have large uncertainties in
their energies and momenta.  Thus, all possibilities for different energies and
momenta are coherently added, taking into account the corresponding phases
accumulated during the propagation from the creation (spacetime) point to the
observation points.

In the following section we will calculate the twin-photon wavefunction by
evaluating the integrals in Eq.~(\ref{scatt}) using reasonable approximations.

\section{Calculating the wave function of the twin photons}\label{sec3}

To calculate the wave function of the twin photons using Eq. (\ref{scatt}), we first
make some assumptions. If the incident pump photon is almost monochromatic, we can
approximately write its position-space wave function from the momentum-space wave
function as in the traditional quantum mechanics of massive particles.\footnote{If
the photon is not monochromatic the change of representation is not so simple, as
discussed in Refs.~\onlinecite{birula94} and \onlinecite{sipe95}.} The result is
\begin{equation}\label{psip1}
        \psi_p(\mb{r}',t')\propto\int d^3k_p\, \phi_p(\mb{k}_p)
\mathrm{e}^{i\mb{k}_p\cdot\mb{r}'-i\omega_p t'},
\end{equation}
where we write the decomposition in terms of the wavevectors $\mb{k}_p$ and angular
frequencies $\omega_p={k}_pc/n_p$ (instead of the momenta $\hbar\mb{k}_p$ and
energies $\hbar\omega_p$) to simplify the notation. 
Decomposing $\mb{k}_p\equiv
\mb{q}_p+\sqrt{k_p^2-q_p^2}\,\mb{\hat{z}}$, with $\mb{q_p}$ being the component of
$\mb{k}_p$ in the $xy$-plane, we note that $\mb{q}_p$ and $\omega_p$ completely
determine $\mb{k}_p$. Thus, we can write the momentum-space wave function in terms
of $\mb{q}_p$ and $\omega_p$, changing the integrals in Eq. (\ref{psip1}) from
$d^3k_p$ to integrals in $d\mb{q}_p$ and $d\omega_p$, giving
\begin{equation}\label{psip}
        \psi_p(\mb{r}',t')\propto\int d\mb{q}_p \int d\omega_p
\phi_p(\mb{q}_p,\omega_p) \mathrm{e}^{i\mb{k}_p\cdot\mb{r}'-i\omega_p t'}.
\end{equation}

This decomposition in terms of $\mb{q}_p$ and $\omega_p$ is useful when the incident
pump photon is in a beam mode that propagates nearly parallel to the $z$-direction,
in the paraxial regime. We will assume that we are operating in the paraxial regime
so that $\phi_p(\mb{q}_p,\omega_p)$ has non-negligible values only for
$q_p\ll k_p$. Writing $\mb{r}'\equiv \bs{\rho}'+z'\mb{\hat{z}}$, where $\bs{\rho}'$
is the component of $\mb{r}'$ in the $xy$ plane, we have
$\mb{k}_p\cdot\mb{r}'=\mb{q}_p\cdot\bs{\rho}'+\sqrt{k_p^2-q_p^2}\,z'$. We will use a
similar notation for the wavevectors of photons 1 and 2. We will also consider that the
nonlinear crystal has $x$- and $y$-dimensions much larger than the width of the beam
mode so that the $x'$ and $y'$ integrals can be extended up to infinity in
Eq.~(\ref{scatt}), and a small dimension in the $z$ direction. Under these approximations, the wave function of
the twin photons from Eq. (\ref{scatt}) can be written as
\begin{eqnarray}\nonumber\label{calc_psi2}
        &&\psi^{(2)}(\mb{r}_1,\mb{r}_2,t)\propto \int d\mb{q}_p \int d\omega_p \int
d\mb{q}_1 \int d\omega_1 \int d\mb{q}_2 \int d\omega_2\\\nonumber
        &&\times\phi_p(\mb{q}_p,\omega_p) \int d\bs{\rho}'
\mathrm{e}^{i(\mb{q}_p-\mb{q}_1-\mb{q}_2)\cdot\bs{\rho}'} \int_{-\infty}^t dt'
\mathrm{e}^{-i(\omega_p-\omega_1-\omega_2)t'}\\
        &&\times\mathrm{e}^{i\mb{k}_1\cdot\mb{r}_1-i\omega_1t}\mathrm{e}^{i\mb{k}_2\cdot\mb{r}_2-i\omega_2t},
\end{eqnarray}
where the integral in $z'$ gives a constant if we also have $k_1\gg q_1$ and $k_2\gg
q_2$. 
In the appendix we discuss the relation
between the omitted integral in $z'$ and the phase matching condition for the
efficient generation of photon pairs. The integral in $\bs{\rho}'$ results in a term
proportional to the 2-dimensional delta function
$\delta^{(2)}(\mb{q}_p-\mb{q}_1-\mb{q_2})$. If the time light takes to propagate
from the crystal to the observation points $\mb{r}_1$ and $\mb{r}_2$ is greater than
the duration of the incident photon pulse, the $t'$ integral can be extended to
$t'=\infty$, resulting in a term proportional to
$\delta(\omega_p-\omega_1-\omega_2)$. So the wave function of the twin photons can
be written as
\begin{eqnarray}\nonumber \label{psi2r}
        &&\psi^{(2)}(\mb{r}_1,\mb{r}_2,t)\propto  \int d\mb{q}_1 \int d\omega_1 \int
d\mb{q}_2 \int d\omega_2\\
        &&\;\;\;\times\;\phi_p(\mb{q}_1+\mb{q}_2,\omega_1+\omega_2) 
        \mathrm{e}^{i\mb{k}_1\cdot\mb{r}_1-i\omega_1t}\mathrm{e}^{i\mb{k}_2\cdot\mb{r}_2-i\omega_2t}.
\end{eqnarray}

We can see that the same decomposition for a one-photon wave function used in Eq.
(\ref{psip}) is used in Eq. (\ref{psi2r}) for a two-photon wave function, such that
the twin-photon wave function in the $(\mb{q},\omega)$ representation is 
\begin{equation}\label{psi2q}
        \phi^{(2)}(\mb{q}_1,\omega_1,\mb{q}_2,\omega_2)\propto\phi_p(\mb{q}_1+\mb{q}_2,\omega_1+\omega_2).
\end{equation}

We can see in the above wave function that the sum of the energies of the twin
photons ($\hbar\omega_1+\hbar\omega_2$) must be equal to the energy of the incident
photon ($\hbar\omega_p$), so energy is conserved in the process. Also, the sum of
the $xy$-momentum of the twin photons ($\hbar\mb{q}_1+\hbar\mb{q}_2$) must be equal
to a $xy$-momentum of the incident photon ($\hbar\mb{q}_p$), so the transverse
momentum is also conserved in the process.\footnote{It is important to stress that
the form $\hbar \mb{k}$ is not the only possibility for the momentum of a photon
with wavevector $\mb{k}$ in a medium, as discussed in Refs.~\onlinecite{pfeifer07}
and \onlinecite{saldanha10}. There are many different ways of dividing the total
momentum of an electromagnetic wave in a linear medium into electromagnetic and
material parts, all of which are compatible with momentum conservation, and for each
chosen division the photon momentum will have a different expression. The form
$\hbar \mb{k}$ is the simplest one for the present case, which is why we have used
it.}
According to the appendix we also have $\hbar k_{1z}+\hbar k_{2z}\approx\hbar
k_{pz}$, where $k_{iz}$ corresponds to the $z$ component of the wavevector of photon
$i$, so that the total momentum is conserved in the process. But note that we did
not impose conservation of energy or momentum in any part of the calculation. Energy
conservation is a consequence of the coherent superposition of the probability
amplitudes of the pair creation at any instant of time, by way of the delta function
$\delta(\omega_p-\omega_1-\omega_2)$ being a consequence of the $t'$ integral in
Eq.~(\ref{calc_psi2}). In the same way, conservation of momentum is a consequence of
the coherent superposition of the probability amplitudes of the pair creation at any
position inside the crystal, by way of the 2-dimensional delta function
$\delta^{(2)}(\mb{q}_p-\mb{q}_1-\mb{q_2})$ being a consequence of the $\bs{\rho}'$
integral in Eq.~(\ref{calc_psi2}). These conservation laws can be obtained simply
from the interference of the probability amplitudes associated with
indistinguishable situations.

\section{Energy and momentum entanglement of the system}\label{sec4}

The twin-photon wave function of Eq.~(\ref{psi2q}) cannot be written as a product of
a wave function for photon 1 and a wave function for photon 2. This condition
characterizes the entanglement of the system. Thus, even if the incident photon has
narrow energy ($\hbar\omega_p$) and momentum ($\hbar\mb{k}_p$) distributions, we see
that the sum of the energies of the generated photons is equal to the energy of the
incident photon, even though the energy of each photon can assume a large range of
values. Similarly, the sum of the momenta of the generated photons is equal to the
momentum of the incident photon, even though the momentum of each photon can also
assume a large range of values. This situation characterizes a state where the twin
photons are highly entangled in both energy and momentum. 

An entangled state of the form of Eq. (\ref{psi2q}) was used by Einstein, Podolsky,
and Rosen (EPR) in their famous paper on the completeness of  quantum
mechanics.\cite{einstein35} Let us consider the wave function for the $x$-component
of the photon momenta when the incident photon is a plane wave propagating in the
$z$-direction. In this case, we can write
$\phi^{(2)}(p_1,p_2)\propto\delta(p_1+p_2)$, with $p_i=\hbar {q}_{ix}$. Of
course, in a realistic situation the delta function must be seen as an approximation
for a very narrow distribution. On the other hand, if we consider the position-space
wave function at the exit face of the crystal (assumed to be extremely thin), we
have $\psi^{(2)}(x_1,x_2)\propto\delta(x_1-x_2)$, because the twin photons are born
at the same position. Again, the delta function must be seen as an approximation.

Consider now that the generated photons reach two observers Alice and Bob who are
placed in arbitrarily separated regions of space. If Bob measures the $x$-momentum
component of his photon and obtains the value $P$, we can state with certainty that
if Alice measures the $x$-momentum component of her photon she will obtain the value
$-P$. So EPR says there is an \textit{element of physical reality} associated with
the momentum of Alice's photon. On the other hand, if Bob measures the $x$-position
of his photon using lenses to project an image of the crystal in the observation
plane and obtains the value $X$, we can state with certainty that if Alice measures
the $x$-position of her photon in a similar way she will also obtain the value $X$,
and EPR says that there is an \textit{element of physical reality} associated with
the position of Alice's photon. However, if Bob is arbitrarily far away from Alice,
his measurement cannot affect her photon in any way, so EPR argues that Alice's
photon should have \textit{elements of physical reality} associated with both
position and momentum. However, in quantum mechanics two observables represented by
noncommuting operators cannot have definite and (in principle) predictable values
simultaneously.  Therefore, EPR argues that quantum mechanics is not a complete
theory because ``hidden variables'' not considered by the theory would be essential
to guarantee the simultaneous reality for the position and momentum of Alice's
photon in the \textit{gedanken} experiment discussed above.

Bohr's reply\cite{bohr35} to the EPR argument was that the measurements of position
or momentum made by Bob are mutually incompatible experiments. Bob's choice of which
experiment he will perform on his photon determines different types of predictions
he can make for experiments made by Alice. And there is no experiment that Alice can
perform on her photon which would reveal the experiment performed by Bob. Bohr
argued that quantum mechanics is indeed a complete theory because it can predict the
probability of the experimental results of any combination of compatible
experiments, which is the actual objective of the theory (such that there are no
paradoxes).

This discussion continued on philosophical grounds until the seminal work of John
Bell.\cite{bell64} Bell showed that there are quantum entangled states with
correlations between two parties that are stronger than what is allowed  by ``hidden
variables'' theories. The fact that experiments have confirmed the quantum
mechanical predictions\cite{aspect82} (although there are still some loopholes) has
changed the way we view the world. We cannot have a consistent local and realistic
description of nature because in an entangled state like Eq.~(\ref{psi2q}), we
cannot attribute reality to each of two separated photons. We must consider either
that Bob's measurement instantaneously changes the state of Alice's photon,
violating locality (but not permitting any instantaneous transmission of information
between Bob and Alice), or that the experimental results of the measurements
performed by Alice do not depend only on the properties of her photon (independently
of Bob's photon) and on her measuring apparatus (independently of Bob's apparatus),
violating any reasonable definition of reality for her photon properties. 
Accessible discussions of these fundamental aspects of quantum mechanics can be
found in chapter 6 of Ref.~\onlinecite{peres} and in Ref. \onlinecite{mermim85}.

A physical implementation of the EPR state was performed by Howell and co-workers
using the twin photons produced by parametric downconversion as described
above.\cite{howell04} The high degree of entanglement in Eq.~(\ref{psi2q}) renders
the twin photons generated by parametric downconversion an extremely valuable
resource for testing such fundamental aspects of quantum mechanics.  Moreover,
entangled states have correlations among parties of the system that are stronger
than what is  allowed by classical physics, and some quantum information protocols
take advantage of these quantum correlations to be more secure or efficient than
their classical counterparts.\cite{nielsen} The twin photons generated by parametric
downconversion have also been extensively used in the implementation of these
quantum information protocols.\cite{nielsen}

\section{Conclusions}\label{sec5}
We derived an expression for the quantum state of the twin photons generated in the
process of parametric downconversion by a scattering procedure that coherently sums
the probability amplitudes for the photon pair to be generated at any position
inside the crystal and at any time. The quantum state we find is equivalent to the
one obtained using the traditional perturbative approach to calculate the
Hamiltonian evolution of an electromagnetic field interacting with a nonlinear
medium.\cite{mandel,hong85,monken98} Our treatment, however, is more intuitive, and
provides useful physical insight to the problem. The twin-photon state is highly
entangled in both energy and momentum, making this system very useful for
experimental tests of fundamental aspects of quantum mechanics and for the
implementation of quantum information protocols. Many of the exciting properties of the system entanglement can be verified in undergraduate laboratories.\cite{dehlinger02,galvez05,carlson06,pearson10,galvez10} A more formal treatment of the work
presented here can be found in Ref.~\onlinecite{saldanha11}.

\acknowledgments

The authors acknowledge J\'ulia E. Parreira for very useful comments on the
manuscript. This work was supported by the Brazilian agencies CNPq, FAPEMIG  and
FACEPE.

\appendix*

\section{Phase matching in parametric downconversion}

In this appendix we discuss the necessary phase matching for the efficient
generation of photon pairs in parametric downconversion.

The phase matching condition can be obtained from the integral in $z'$ omitted from
Eq.~(\ref{calc_psi2}). Approximating $\sqrt{k_i^2-q_i^2}\approx k_i-q_i^2/(2k_i)$,
we have
\begin{equation}\label{ap}
        I\propto\int_{-L/2}^{L/2} dz'
\,\mathrm{e}^{i(k_p-k_1-k_2)z'}\,\mathrm{e}^{-i(q_p^2/k_p-q_1^2/k_1-q_2^2/k_2)z'/2},
\end{equation}
where the width of the crystal in the $z$ direction is $L$. The efficiency of
photon-pair generation is proportional to $I$, so we must have $k_p=k_1+k_2$ to
obtain collinear generation, in which the generated photons propagate close to the
$z$-axis; otherwise the oscillations of the first exponential above with $z'$ will
decrease the value of $I$. The wavenumber of each photon can be written as
$k_i(\omega_i)=\omega_i n_i(\omega_i)/c$, where $n_i(\omega_i)$ is the refractive
index for photon $i$ with angular frequency $\omega_i$. Let us assume that each of
the  photons is post-selected in a relatively narrow range of frequencies such that
the refractive index for each photon can be considered a constant $n_i$; this can be
accomplished with the use of interference filters in the photon detectors that are
used to detect the twin photons. Because Eq.~(\ref{psi2q}) states that the sum of
the frequencies of the generated photons is equal to the frequency of the incident
photon, the phase matching condition for collinear generation can be written as
$n_p=(n_1+n_2)/2$. However, the inherent dispersion of the material makes the
refractive index increase with frequency, so that in an isotropic medium we have
$n_p>(n_1+n_2)/2$, thus making it impossible to efficiently generate photon pairs.
This problem is solved with the use of a nonlinear and birefringent material such
that the refractive index also depends on the polarization of the photons, giving
$n_p=(n_1+n_2)/2$ for photons with different polarization states. Type-I parametric
downconversion corresponds to the case in which both generated photons have
polarization orthogonal to the polarization of the incident photon, and type-II
parametric downconversion has one of the generated photons with polarization
orthogonal to, and the other one parallel to, the incident photon.

In cases where $k_p-k_1-k_2=-\alpha<0$, we can have an efficient generation of
photon pairs that propagate in directions making angles with the $z$-axis such that
$q_1^2/k_1+q_2^2/k_2=2\alpha$ and the second exponential in Eq.~(\ref{ap}) cancels
the first ($q_p^2/k_p$ is assumed to be negligible). For this reason the generated
photons are usually emitted around cones. In any case we have $k_{1z}+k_{2z}\approx
k_{pz}$, where $k_{iz}$ corresponds to the $z$-component of the wavevector of photon
$i$.

Another important approximation that we make in our treatment is to consider that
the integrals in $d\mb{q}_1$ and $d\mb{q}_2$ in Eq. (\ref{calc_psi2}) can be computed
without considering the dependence of $I$ on $\mb{q}_1$ and $\mb{q}_2$ in Eq.
(\ref{ap}). This is a good approximation for crystals with small width $L$, because
the accumulated phase in the second exponential in  Eq. (\ref{ap}) becomes
negligible once the phase matching condition is established.


\end{document}